\documentclass[]{mn2e}
\usepackage{graphicx}
\usepackage{epsfig}
\usepackage{amsmath}
\newcommand       \Angstrom     {\,{\rm \AA}}

\newcommand       \cm           {\,{\rm cm}}

\newcommand       \erg          {\,{\rm erg}}

\newcommand       \g            {\,{\rm g}}

\newcommand       \s            {\,{\rm s}}

\newcommand     \gtsim  {\lower.5ex\hbox{$\buildrel > \over \sim$}}
\newcommand     \ltsim  {\lower.5ex\hbox{$\buildrel < \over \sim$}}
\newcommand     \simgt  {\lower.5ex\hbox{$\buildrel > \over \sim$}}
\newcommand     \simlt  {\lower.5ex\hbox{$\buildrel < \over \sim$}}

\newcommand       \mum          {\,{\rm \mu m}}

\newcommand       \Msun         {\,{\rm M}_\odot}

\newcommand       \mH           {m_{\rm H}}
\newcommand       \simali       {\sim\,}
\newcommand       \magni        {\,{\rm mag}}

\newcommand       \NH          {N_{\rm H}}
\newcommand       \NC          {N_{\rm C}}

\newcommand       \CTOHgraa  {\left[{\rm C/H}\right]_{\rm gra}}
\newcommand       \CTOHgrab  {\left[{\rm \frac{C}{H}}\right]_{\rm gra}}
\newcommand       \rhogra  {\rho_{\rm gra}}
\newcommand       \etagra  {\eta_{\rm gra}}
\newcommand       \Fobs  {F_\lambda^{\rm obs}}
\newcommand       \Fintr  {F_\lambda^{\rm intr}}
\newcommand       \Fmod  {F_\lambda^{\rm mod}}
\newcommand       \taulambda  {\tau_\lambda}
\newcommand       \Cext  {C_{\rm ext}}
\newcommand       \Ndata  {N_{\rm data}}
\newcommand       \sigobs  {\sigma_{{\rm obs},\lambda}}
\newcommand       \sigobsp  {\sigma^{\prime}_{{\rm obs},\lambda}}
\newcommand       \omegap  {\omega^{\prime}}
\newcommand       \widthWL  {\Delta\lambda}
\newcommand       \widthWV  {\gamma}
\title{
What causes the ultraviolet extinction bump at the cosmic dawn?
}
\author[Li, Yang \& Li]
            {Qi~Li$^{1,2}$,
              X.J.~Yang$^{1,2}$\thanks{xjyang@xtu.edu.cn},
             and Aigen Li$^{2}$\thanks{lia@missouri.edu}\\
 $^1$Hunan Key Laboratory for Stellar
              and Interstellar Physics
              and School of Physics and Optoelectronics,
              Xiangtan University, Hunan 411105, China\\
  $^2$Department of Physics and Astronomy,
                  University of Missouri,
                  Columbia, MO 65211, USA\\
                  }

\begin{document}

\date{}
\pagerange{\pageref{firstpage}--\pageref{lastpage}} \pubyear{2024}

\maketitle

\label{firstpage}
\begin{abstract}
The enigmatic ultraviolet (UV) extinction bump
at 2175$\Angstrom$, the strongest spectroscopic
absorption feature superimposed on the interstellar
extinction curve, has recently been detected
at the cosmic dawn
by the James Webb Space Telescope (JWST)
in JADES-GS-z6-0, a distant galaxy at redshift
$z\approx6.71$,
corresponding to a cosmic age of just 800
million years after the Big Bang.
Although small graphite grains have historically
long been suggested as the carrier of
the 2175$\Angstrom$ extinction bump
and graphite grains are expected to have already
been pervasive in the early Universe,
in this work we demonstrate that small graphite
grains are not responsible for the UV extinction bump
seen at the cosmic dawn in JADES-GS-z6-0,
as the extinction bump arising from
small graphite grains is too broad and peaks
at wavelengths that are too short to be consistent
with what is seen in JADES-GS-z6-0.
\end{abstract}

\section{Introduction}\label{sec:intro}
Due to its unprecedentedly high sensitivity,
the {\it James Webb Space Telescope} (JWST)
is capable of detecting rest-frame
ultraviolet (UV) emission
from very high-redshift galaxies,
and therefore provides
a paradigm-shifting view
of early galaxy evolution.
The rest-frame UV starlight emission
is absorbed and scattered by solid dust grains
in the galaxy's interstellar medium (ISM).
The wavelength dependence of
the dust absorption and scattering---their
combination is called extinction---exhibits
a strong spectral band peaking at
$\simali$2175$\Angstrom$.
As the most prominent spectroscopic
feature in the interstellar extinction curve,
the 2175$\Angstrom$ extinction bump
is widely seen in the Milky Way
(e.g., see Valencic et al.\ 2004)
and external galaxies, both near
(e.g., see Decleir et al.\ 2019,
Wang et al.\ 2022, Gordon et al.\ 2024)
and far (see Shivaei et al.\ 2022, Lin et al.\ 2023,
and references therein).

More recently, utilizing the {\it Near Infrared
Spectrograph} (NIRSpec) on board JWST,
Witstok et al.\ (2023) detected
the 2175$\Angstrom$ extinction bump
in  JADES-GS+53.15138-27.81917
(also known as JADES-GS-z6-0),
a distant galaxy at redshift $z\approx6.71$,
corresponding to a cosmic age of just 800
million years (Myr) after the Big Bang.
%

The identification of the carrier of
the 2175$\Angstrom$ extinction bump
is relevant for interpreting the UV emission
from distant galaxies observed with JWST.
Immediately after its detection
nearly 60 years ago (Stecher 1965),
the 2175$\Angstrom$ bump was
attributed to small graphite grains
(Stecher \& Donn 1965).
If the bump is indeed due to small graphite grains,
the detection of the 2175$\Angstrom$ extinction
bump at  $z\approx6.71$ indicates that galaxies
less than 1\,Gyr from the Big Bang were already
forming abundant small graphite grains.
To interpret JWST observations of
high redshift galaxies,
it is imperative to place constraints
on the nature of the 2175$\Angstrom$
bump carrier.
In this regard, JADES-GS-z6-0 provides
a valuable opportunity to examine this
enigmatic extinction bump.
In this work, we aim to quantitatively
investigate if small graphite grains
can explain the UV extinction bump
detected by JWST in JADES-GS-z6-0.
This paper is organized as follows.
In \S\ref{sec:model} we model the observed
UV extinction bump in terms of small graphite grains.
The results and implications are discussed
in \S\ref{sec:discussion}.
Our major conclusion is summarized
in \S\ref{sec:summary}.

\begin{figure*}
\begin{center}
\includegraphics[width=10cm,angle=0]{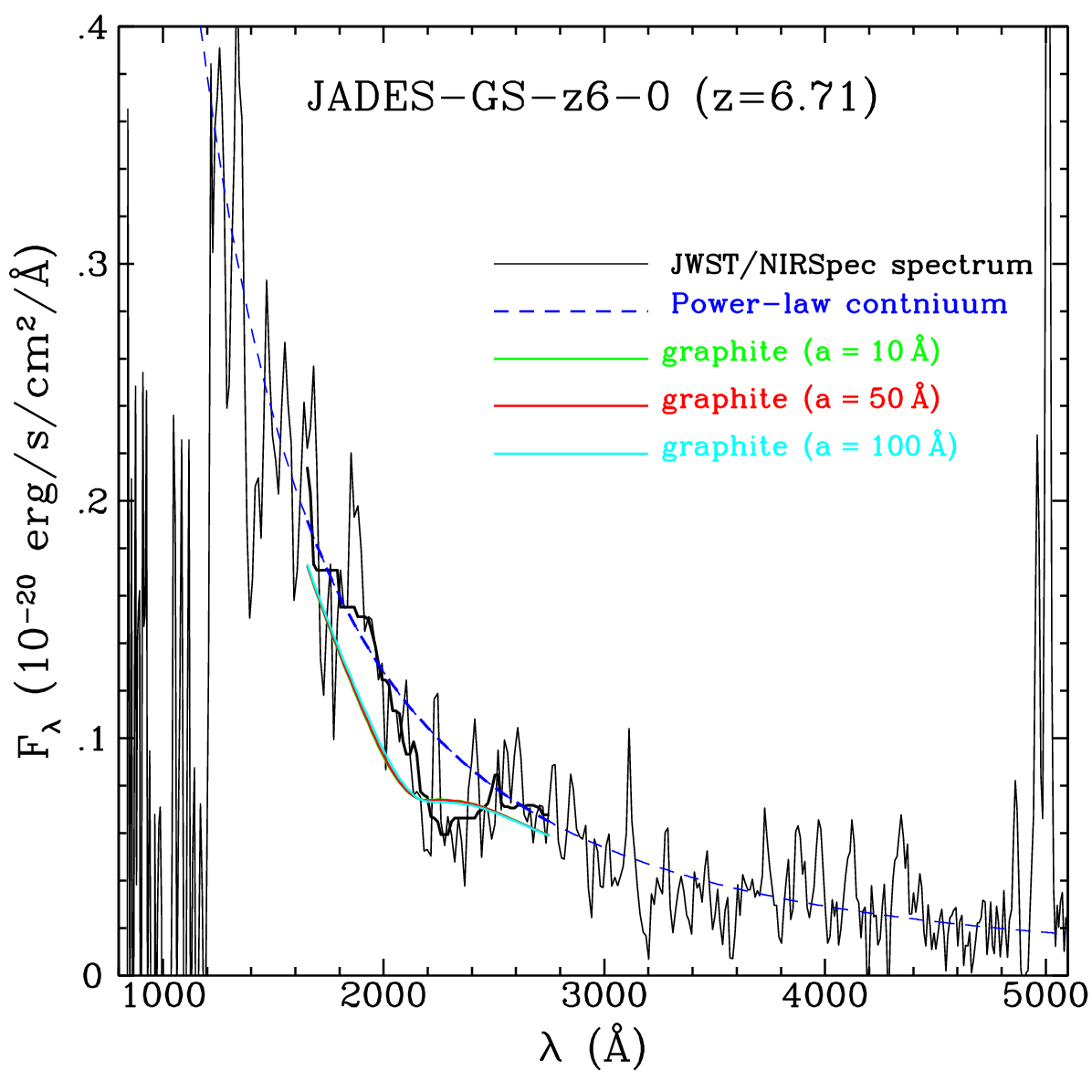}
\end{center}
\caption{\label{fig:sed}
  Comparison of the rest-frame UV spectrum
  of JADES-GS-z6
  measured by JWST/NIRSpec (gray solid line)
  with the dust-attenuated model spectra
  generated from the power-law continuum
  (blue dashed line) attenuated by extinction
  arising from spherical graphite grains
  of radii of $a$\,=\,10$\Angstrom$ (green solid line),
  50$\Angstrom$ (red solid line), and
  100$\Angstrom$ (cyan solid line).
  Note that the green solid line
  is essentially hidden by the red solid line.
  The thick black solid line represents
  a running median and highlights the UV bump region
  around 2175$\Angstrom$.
  Apparently, graphite dust produces
  too broad a UV bump to agree with
  the JWST/NIRSpec spectrum.
  The model UV bump also fails to peak
  at the same wavelength as observed by JWST/NIRSpec.
        }
\end{figure*}

\section{Graphite as An Explanation
            for the Extinction Bump}\label{sec:model}
JADES-GS-z6 is located in a portion of
an area of the sky known as GOODS-South,
which was observed as part of
the {\it JWST Advanced Deep
Extragalactic Survey} (JADES;
see Eisenstein et al.\ 2023).
JADES-GS-z6 is a low-metallicity,
normal star-forming galaxy
with a subsolar metallicity
of $Z$\,$\approx$\,0.2--0.3\,$Z_{\odot}$
and a star formation rate (SFR)
of a few solar masses per year
(Witstok et al.\ 2023).
The Balmer decrement
as measured from the H$\alpha$/H$\beta$
line ratio indicates a nebular extinction
of $E(B-V)\approx0.25\pm0.07\magni$
(see Witstok et al.\ 2023),
demonstrating that this galaxy suffers
significant dust obscuration.
As illustrated in Figure~\ref{fig:sed},
the rest-frame UV spectrum obtained by JWST
exhibits a pronounced dip at wavelength
$\lambda_0$\,$\simali$2263$^{+20}_{-24}\Angstrom$,
with a width (measured in wavelength) of
$\Delta\lambda$\,$\simali$250$\Angstrom$
which corresponds to width
(measured in inverse wavelength)
$\widthWV$\,$\simali$0.49$\mum^{-1}$.\footnote{%
Witstok et al.\ (2023) determined these
         parameters by fitting the dip with
         a Drude profile of width $\Delta\lambda$
         centered at $\lambda_0$.
       }
This dip resembles the interstellar UV extinction
bump but shifting the nominal peak wavelength
of 2175$\Angstrom$ to a somewhat longer
wavelength. Its width is considerably narrower
than the mean width of
$\widthWL\simali$473$\Angstrom$
(or $\widthWV\simali$1$\mum^{-1}$)
of the Milky Way ISM (Valencic et al.\ 2004).
In the following, we will follow the custom and
refer this dip as the ``2175$\Angstrom$'' extinction
bump despite it peaks at 2263$\Angstrom$.
%

\begin{figure*}
\begin{center}
\hspace{-1cm}
\begin{minipage}[t]{0.4\textwidth}
\resizebox{8.5cm}{7.5cm}{\includegraphics[clip]{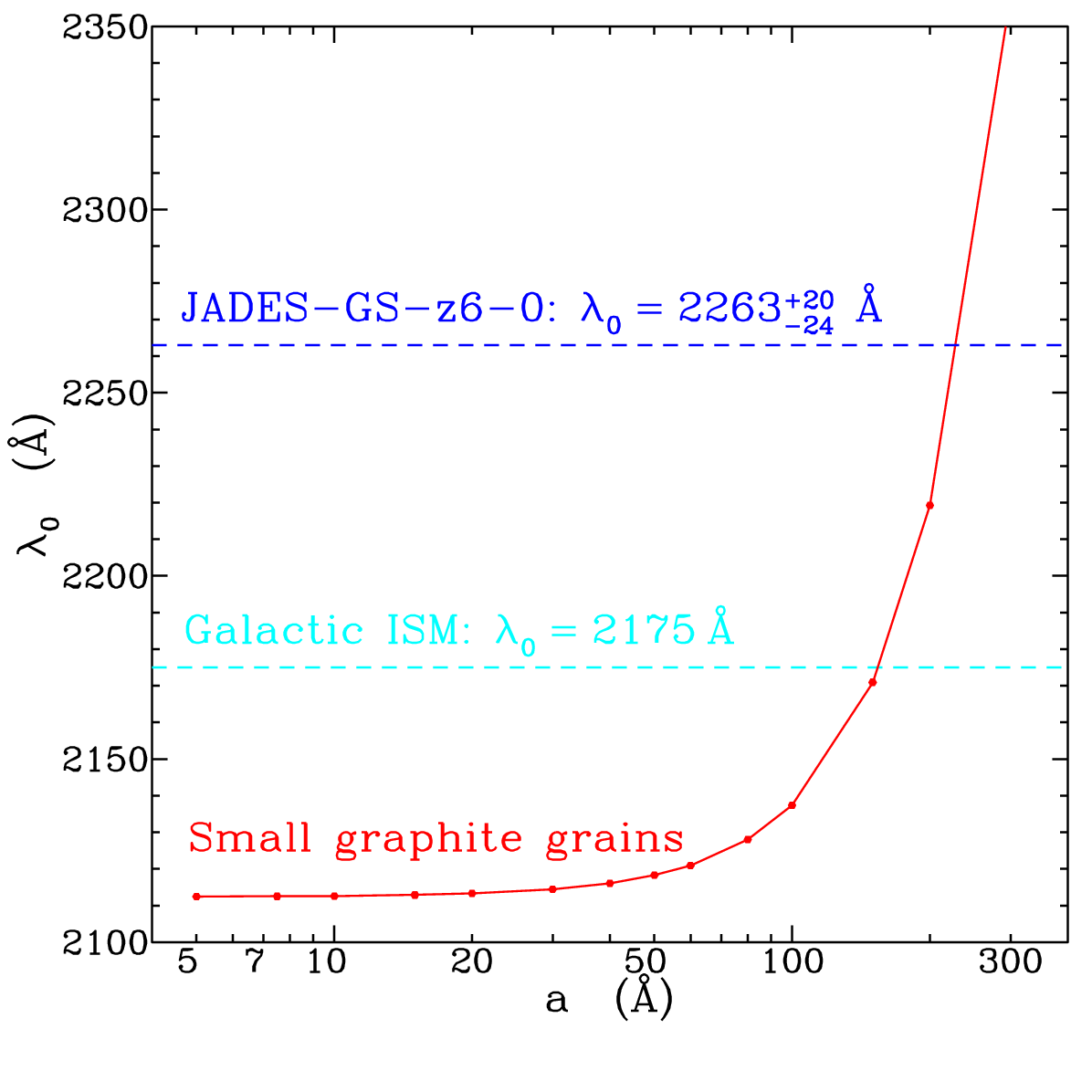}}\vspace{-0.5cm}
\end{minipage}
\hspace{1cm}
\begin{minipage}[t]{0.4\textwidth}
\resizebox{8.5cm}{7.5cm}{\includegraphics[clip]{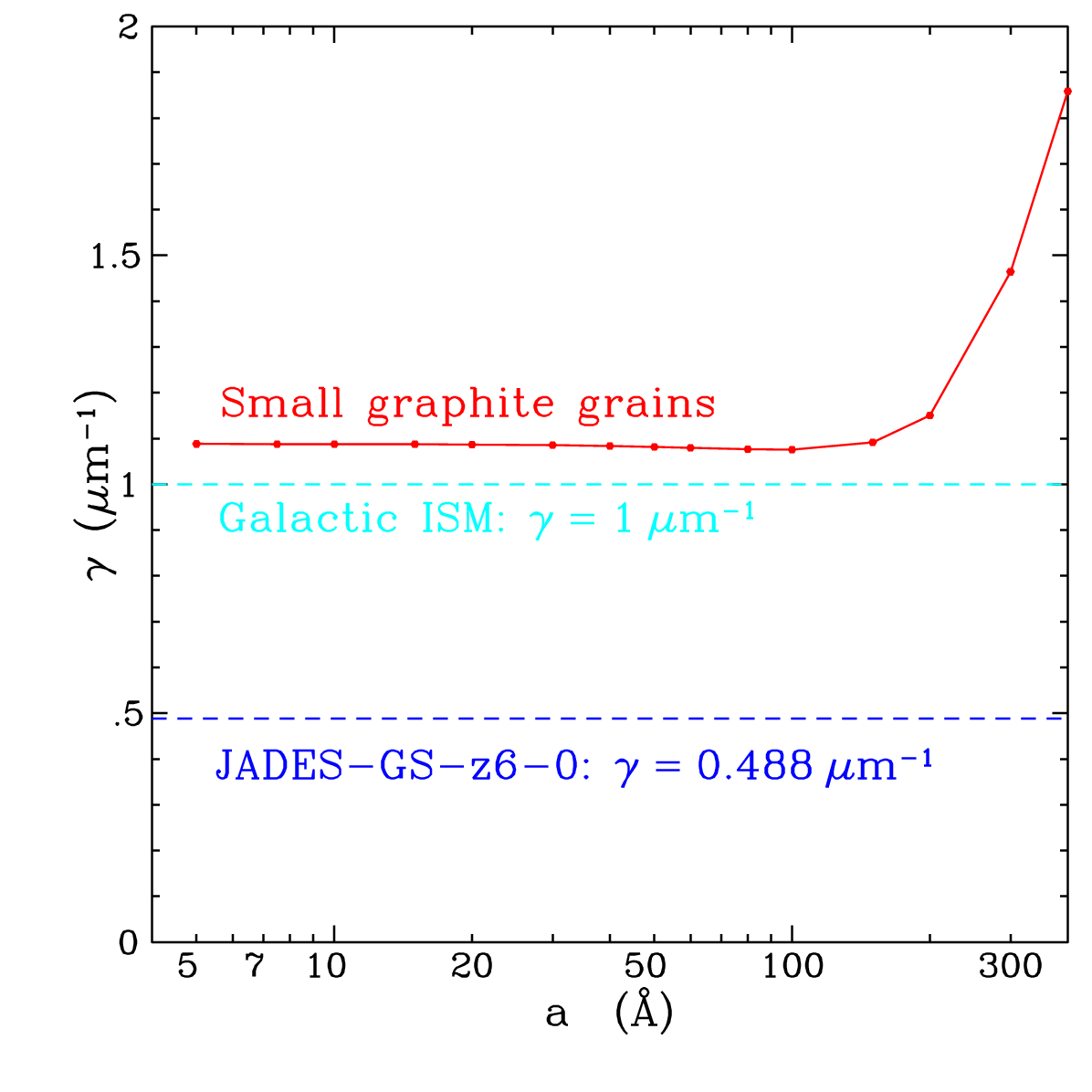}}\vspace{-0.5cm}
\end{minipage}
\end{center}
\caption{
         \label{fig:widthpeak}
         Central wavelength ($\lambda_0$; left panel)
         and width ($\widthWV$; right panel)
         of the UV bump calculated from small graphite spheres
         of radii ($a$) ranging from a few angstroms to a few tens of
         nanometers. Horizontal lines plot the central wavelengths
         and widths of the Galactic average extinction bump
         and that of the high-$z$ galaxy JADES-GS-z6
         at an epoch of the cosmic dawn.
         It is apparent that the UV bumps resulting from
         small graphite grains are far too broad and
         peak at wavelengths that are too short
         to be consistent with that observed in JADES-GS-z6.
         }
\end{figure*}

Let $\Fintr$ be the intrinsic, ``extinction-free‘’
flux emitted by JADES-GS-z6 at wavelength
$\lambda$ that would be ``detected''
by JWST,\footnote{%
    Strictly speaking, this is not dust-free,
    instead, it is a {\it continuum} emission
    which suffers continuum dust attenuation
    but just free of {\it extra} extinction at the bump.
    For more details see \S\ref{sec:discussion}.
  }
and $\Fmod$ be the dust-attenuated ``model'' flux
expected to be observed by JWST.
$\Fmod$ and $\Fintr$ are related through
\begin{equation}\label{eq:FobsFintr}
\Fmod = \Fintr \exp\left(-\taulambda\right)~~,
\end{equation}
where $\taulambda$ is the optical depth
at wavelength $\lambda$
arising from the dust along the line of sight
toward the region in JADES-GS-z6 observed
by JWST.\footnote{%
    We note that, strictly speaking,
    $\taulambda$ includes also foreground dust
    in the Milky Way galaxy along the line of sight
    toward JADES-GS-z6. However, the foreground
    reddening between $g$ and $r$ bands
    derived from the Galactic extinction map
    of Green et al.\ (2015) is only
    $E(g-r)\approx0.025\magni$,
    corresponding to a visual extinction
    of $A_V\approx0.057\magni$.
    Based on the typical wavelength dependence
    of the Galactic UV, optical and near infrared
    extinction (see Cardelli et al.\ 1989,
    Gordon et al.\ 2009, Fitzpatrick et al.\ 2019,
    Gordon et al.\ 2021, Decleir et al.\ 2022,
    Gordon et al.\ 2023, Wang \& Chen 2024),
    this translates into a negligible amount
    of foreground extinction in the NIRSpec
    wavelength range.
    }
Assuming a single dust species of
a single size $a$ and a column density of
$N_d$, we derive the optical depth from
\begin{equation}\label{eq:tau1}
\tau_\lambda = N_d\times\Cext(a,\lambda) ~~,
\end{equation}
where $\Cext(a,\lambda)$ is the extinction
cross section of the dust of size $a$
at wavelength $\lambda$.
Let's consider spherical graphite dust
of radius $a$. Let $\CTOHgraa$ be the amount
of carbon (C) relative to hydrogen (H)
locked up in graphite dust. The dust
column density ($N_d$) relates to
the hydrogen column density ($\NH$) through
\begin{equation}\label{eq:N_d}
N_d = \NH\times
\CTOHgrab\times\frac{12\,\mH}
{\left(4\pi/3\right)\,a^3\rhogra} ~~,
\end{equation}
where $\mH$ is the mass of a hydrogen atom,
and $\rhogra\approx2.22\g\cm^{-3}$
is the mass density of graphite.
Therefore, the optical depth
$\taulambda$ becomes
\begin{equation}\label{eq:tau2}
\taulambda = \NH\times
\CTOHgrab\times\frac{\Cext(a,\lambda)}{\NC} ~~,
\end{equation}
where $\NC$ is the number of C atoms
contained in a graphite dust grain
of radius $a$
\begin{equation}\label{eq:NC}
\NC = \frac{\left(4\pi/3\right)\,a^3\rhogra}{12\,\mH} ~~.
\end{equation}

To best reproduce $\Fobs$, the flux observed by JWST,
we require to minimize
\begin{equation}\label{eq:chi2}
\chi^2 = \sum_{i=1}^{\Ndata}
{\omegap}^2\,\left\{\frac{\ln\Fobs-\ln\Fmod}
{\sigobsp}\right\}^2 ~~,
\end{equation}
where $\Ndata=33$ is the number of
the JWST/NIRSpec data points
of JADES-GS-z6 in the wavelength range
of 1993.1--2488.7$\Angstrom$
where the observed UV extinction bump peaks,
$\omegap$ is the weight we introduce here
to enhance the fit to the bump,
and $\sigobsp$ is the reduced observational
uncertainty:
\begin{equation}\label{eq:sigmaobsp}
\sigobsp \equiv \sigobs/\Fobs ~~,
\end{equation}
where $\sigobs$ is the observational uncertainty.
We adopt the following weight
\begin{equation}\label{eq:weight}
\omegap = \left\{\begin{array}{lr}
0, & |\lambda - \lambda_0| > 2\Delta\lambda,\\
1, & \Delta\lambda < |\lambda - \lambda_0|
        < 2\Delta\lambda,\\
2, & \Delta\lambda/2 < |\lambda - \lambda_0|
        < \Delta\lambda,\\
4, & |\lambda - \lambda_0| < \Delta\lambda/2,\\
\end{array}\right.
\end{equation}
where $\lambda_0$ and $\Delta\lambda$ are
the peak wavelength and full width half maximum
(FWHM) of the ``2175$\Angstrom$'' UV extinction bump.
While rather arbitrarily, we design such a weight
to ``force'' the model spectrum to fit the bump,
with {\it increasing} emphasis paid to the data in
the wavelength range {\it closer} to the bump peak.

By defining $\etagra\equiv\NH\times
\left[{\rm C/H}\right]_{\rm gra}$,
we achieve the best-fit under the condition of
\begin{equation}\label{eq:chi2gra}
\partial{\chi^2}/\partial{\etagra} = 0 ~~,
\end{equation}
and from eq.\,\ref{eq:chi2gra} we derive
\begin{eqnarray}\label{eq:etagra}
 \etagra & = & \sum_{j=1}^{\Ndata}
\left\{
\frac{
\left(\ln\Fintr-\ln\Fobs\right)
\times\left(\Cext/\NC\right)}
{{\sigobsp}^2}
\right\} \nonumber \\
& \times & \left\{
\sum_{j=1}^{\Ndata}
\left[\frac{\Cext/\NC}{\sigobsp}
\right]^2
\right\}^{-1} ~~.
\end{eqnarray}

We use Mie theory to calculate the UV extinction
cross sections of graphite grains
(Bohren \& Huffman 1983).
We adopt the dielectric functions of
graphite of Draine \& Lee (1984)
and Li \& Draine (2001).
Graphite is a highly anisotropic material.
Its dielectric function depends on the orientation
of the electric field relative to the ``crystal axis'',
or ``c-axis'' normal to the ``basal plane''.
For electric fields $\vec{E}$ parallel to the c-axis,
the dielectric function is $\varepsilon_{\|}$;
for $\vec{E}$ perpendicular to the c-axis,
the dielectric function is $\varepsilon_{\bot}$.
%
We take the ``1/3--2/3'' approximation
in which the optical properties of graphite
are represented by a mixture of ``isotropic'' spheres,
with 1/3 of which have dielectric functions
$\varepsilon = \varepsilon_{\|}$;
and 2/3 of which have
$\varepsilon = \varepsilon_{\bot}$
(see Draine \& Malhotra 1993).

We calculate the extinction cross sections
of graphite in the wavelength range of
0.1--0.5$\mum$. We consider a range
of graphite sizes, from $a$\,=\,5$\Angstrom$
up to  $a$\,=\,0.3$\mum$.
Both wavelengths and sizes are finely spaced.
We present in Figure~\ref{fig:sed} the best-fit
model spectra given by several graphite sizes.
For illustrative purposes, we show the model
spectra $\Fmod(\lambda)$ created by attenuating
the intrinsic spectrum $\Fintr(\lambda)$
with an optical depth $\taulambda$
arising from spherical graphite dust grains
of radii of $a$\,=\,10$\Angstrom$,
50$\Angstrom$ and 100$\Angstrom$.
Following Witstok et al.\ (2023),
we approximate the intrinsic spectrum
by a power-law spectrum
which fits the UV continuum
observed by JWST/NIRSpec:\footnote{%
    As mentioned earlier,
    this power-law UV ``continuum''
    spectrum is not really the true,
    dust-free emission from JADES-GS-z6.
    It merely approximates
    the intrinsic continuum emission
    of JADES-GS-z6 subjected to
    continuum dust attenuation.
  }
\begin{equation}\label{eq:continuum}
\Fintr(\lambda) = \left(0.236\pm0.012\right)
\times10^{-20}\times\left(\frac{\lambda}
{1500\Angstrom}
\right)^{-2.13}\erg\s^{-1}\cm^{-2}\Angstrom^{-1}~~.
\end{equation}
We find that no matter what sizes are assumed
for graphite dust, the model spectra
show a bump too broad to
reproduce the observed UV bump.

\section{Discussion}\label{sec:discussion}
In the Milky Way, the central wavelength
($\lambda_0$) of the 2175$\Angstrom$
extinction bump is essentially invariant,
while the bump width ($\widthWL$) shows
considerable variation: $\lambda_0$ varies
by only $\pm$0.46\% (2$\sigma$)
around 2175$\Angstrom$,
in contrast, $\widthWV$ varies by
$\pm$12\% (2$\sigma$) around 473$\Angstrom$
(see Fitzpatrick \& Massa 1986,
Valencic et al.\ 2004, Wang et al.\ 2023).
With $\lambda_0\approx2263^{+20}_{-24}\Angstrom$,
the UV bump seen in JADES-GS-z6 peaks at
a substantially longer wavelength.
JADES-GS-z6 exhibits a much narrower bump
($\widthWL\approx250\Angstrom$,
$\widthWV\approx0.49\mum^{-1}$),
even narrower than the narrowest UV bumps
of the Milky Way interstellar lines of sight
($\widthWV\approx0.75\mum^{-1}$;
see Valencic et al.\ 2004).

In Figure~\ref{fig:widthpeak} we plot
the central wavelength $\lambda_0$
and width $\widthWV$ of the UV bump
arising from small graphite spheres
as a function of radius $a$.
It clearly demonstrates that small graphite
grains (with $a<230\Angstrom$) produce
a UV bump peaking at a wavelength
too short to agree with what was seen
in JADES-GS-z6.
Also, the UV bump caused by small
graphite spheres is far too broad
to explain that of JADES-GS-z6.
Complementary to Figure~\ref{fig:sed},
Figure~\ref{fig:widthpeak} clearly demonstrates
that small graphite spheres are not able
to account for the UV bump detected
at the cosmic dawn in JADES-GS-z6.
We note that for $a\simlt100\Angstrom$
the bump central wavelength and width
show little variations with grain size.
This is not unexpected: for grains
in the Rayleigh regime ($2\pi\,a/\lambda\ll 1$),
the extinction profile is essentially independent
of grain size (see Bohren \& Huffman 1983).

So far, we have focused on spherical grains.
Interstellar grains must be nonspherical
as indicated by the detection of interstellar
polarization. However, for most sightlines,
the 2175$\Angstrom$ extinction bump
is unpolarized.\footnote{%
  To date, only two Galactic lines of sight
  toward HD\,147933 and HD\,197770
  have a weak 2175$\Angstrom$ polarization
  feature detected (Clayton et al.\ 1992;
  Anderson et al.\ 1996).
  }
The lack of 2175$\Angstrom$ polarization
does not necessarily imply that the carriers
of the 2175$\Angstrom$ extinction bump
must be spherical; instead, they could be
nonspherical but not aligned.
However, even if they are nonspherical,
it is still unlikely for them to be capable
of explaining the UV bump
observed in JADES-GS-z6.
As demonstrated in Figure~11 of
Draine \& Malhotra (1993), the bump
increasingly broadens and the bump
peak shifts to shorter wavelengths
as prolate grains become more elongated.
Compared with the bump of JADES-GS-z6,
the UV bump arising from prolate
graphite grains is too broad and peaks
at wavelengths that are too short.
Therefore, prolate graphite grains are
not able to account for the bump of
JADES-GS-z6. For oblate graphite grains,
the bump width decreases
and the bump peak shifts to
longer wavelength as the oblateness
increases. Nevertheless, to produce
a bump as narrow as that of JADES-GS-z6
($\gamma\approx0.49\mum^{-1}$),
one requires extremely flattened oblates
(see the middle panel of Figure~11
of Draine \& Malhotra [1993]).
However, the bump arising from
such extremely elongated oblates
would peak at wavelengths considerably
longer than that of JADES-GS-z6
($\lambda_0^{-1}\approx4.42\mum^{-1}$;
see the upper panel of Figure~11
of Draine \& Malhotra [1993]).
Therefore, we conclude that spheroidal
graphite grains are not able to explain
the UV bump seen in JADES-GS-z6.

As described in \S\ref{sec:model},
the intrinsic spectrum ($\Fintr$) of  JADES-GS-z6
is approximated by the power-law fit to the UV
continuum. The underlying assumption for
this approximation is that there is little extinction
in the wavelength range outside of the UV bump.
This is not true: both in the Milky Way and in nearby
galaxies (e.g., the Large and Small Magellanic Clouds,
and M31), the extinction steeply increases with inverse
wavelength ($\lambda^{-1}$) toward the far-UV.
Therefore, the continuum beyond $\lambda<2000\Angstrom$
is not really the true stellar continuum emission,
instead, it has also been affected by dust extinction.
As a matter of fact, the emission of JADES-GS-z6
at all wavelengths is affected by dust attenuation,
although the attenuation is less severe
at longer wavelengths.
In the wavelength range of the bump,
there is also continuum attenuation.
Nevertheless, in this work as we are mostly
interested in the UV extinction bump, it is therefore
reasonable to ignore the continuum extinction
outside of the wavelength range of the bump.
If we are to derive the extinction curve from
the optical to the far-UV, the power-law fit to
the UV continuum is not a valid representation
of the stellar continuum.

Finally, we note that so far we have treated
the UV bump seen in JADES-GS-z6 as
an extinction feature. However, in extended
objects such as external galaxies,
where individual stars cannot be resolved,
one cannot measure extinction but attenuation.
While extinction represents the loss of starlight
out of the observer's line of sight as a result of both
absorption and scattering, attenuation is a combination
of absorption, scattering out of the line of sight as well
as scattering into the line of sight.
Therefore, the complex geometry between stars and dust
has to be taken into account when one tries to constrain
the nature of the carrier of the attenuation bump.
This requires complex radiative transfer simulations
(e.g., see Seon \& Draine 2016, Salim \& Narayanan 2020).
As radiative transfer would flatten the bump
(e.g., see Seon \& Draine 2016),\footnote{%
  Indeed,  the attenuation curves derived for starburst
  galaxies lack the UV bump at 2175$\Angstrom$
  (Calzetti et al.\ 1994). This is commonly explained
  in terms of radiative transfer effects
  (e.g., see Gordon et al.\ 1997, Witt \& Gordon 2000,
  Fischera et al.\ 2003, Seon \& Draine 2016).
  }
the UV attenuation bump seen in JADES-GS-z6
would imply an even {\it narrower} ``intrinsic''
extinction bump, too narrow to be accounted for
by graphite grains.

We also note that, although small graphite grains
are unlikely the carrier candidate material responsible
for the UV extinction seen in JADES-GS-z6,
graphite is expected to be present
in this galaxy (Yang \& Li 2023).
If star formation began when the Universe was
about 400 Myr old, at the epoch of JADES-GS-z6-0
(about 800 Myr old),
even the oldest stars were only 400 Myr old.
Therefore, there was not enough time for low- to
intermediate-mass stars ($\simali$0.5--8$\Msun$)
to evolve sufficiently to the dust-producing
``asymptotic giant branch'' (AGB) phase.
In the Milky Way, AGB stars dominate
the production of stardust and typically,
it takes them one billion years (Gyr) to evolve.
As only massive stars
(with a stellar mass of at least 8$\Msun$)
would have been able to evolve in such a short time scale
(as that available for JADES-GS-z6-0),
it is often suggested that supernovae were responsible
for the dust in the first billion years of cosmic time
(e.g., see Todini \& Ferrara 2001, Nozawa et al.\ 2003,
Sugerman et al.\ 2006, Bianchi \& Schneider 2007,
Sarangi \& Cherchneff 2015, Yang \& Li 2023).
Supernovae in the Milky Way certainly make graphite
as supernova-condensed graphite grains have been
identified in primitive meteorites (Nittler \& Ciesla 2016).
As supernovae are likely the only dust source
at the cosmic dawn, it is reasonable to expect
graphite grains to be present in the ISM
of JADES-GS-z6-0. Perhaps these graphite grains
in  JADES-GS-z6-0 are like those found in meteorites
that are so large (1--20$\mum$ in diameter) that they
could not produce any absorption band
around 2,175$\Angstrom$.
Therefore, species other than graphite,
e.g., polycyclic aromatic hydrocarbon molecules
(Lin et al.\ 2025), carbon buckyonions
(Li et al.\ 2008), and even T-carbon
(Sheng et al.\ 2011, Ma et al.\ 2020),
must be responsible for the 2175$\Angstrom$ bump.

\vspace{-3mm}
\section{Summary}\label{sec:summary}
We have examined the UV extinction bump
detected by JWST/NIRSpec at an epoch of
cosmic dawn in JADES-GS-z6-0,
a highly-redshifted galaxy at $z\approx6.71$.
It is found that small graphite grains
produce an UV extinction bump
that is too broad and peaks at wavelengths
that are too short to be consistent with that
seen in JADES-GS-z6-0.
Species other than graphite
(e.g., polycyclic aromatic hydrocarbon molecules)
should be explored as possible candidate carriers
for the UV extinction bump.

\vspace{-4mm}
\section*{Acknowledgements}
We thank the anonymous referee
for his/her valuable comments and suggestions.
We thank J.~Witstok for providing us
with the JWST/NIRSpec data of JADES-GS-z6.
We thank B.T.~Draine, Q.~Lin, C.E.~Mentzer,
Q.~Wang, J.~Witstok and B.~Yang
for stimulating discussions.
QL and XJY are supported in part by
NSFC 12122302 and 11873041.

\vspace{-4mm}
\section*{Data Availability}
The data underlying this article will be shared
on reasonable request to the corresponding authors.

\vspace{-4mm}

%

%

%
\end{document}